\documentclass{scrartcl}%,showpacs
\usepackage[T1]{fontenc}
\usepackage[utf8]{inputenc}
\usepackage{mathtools}
\usepackage{amsmath}
\usepackage{amssymb}
\usepackage{graphicx}
\usepackage[unicode=true,pdfusetitle,
 bookmarks=true,bookmarksnumbered=false,bookmarksopen=false,
 breaklinks=false,pdfborder={0 0 1},backref=false,colorlinks=true]
 {hyperref}

\DeclareMathOperator{\linhull}{span}
\DeclareMathOperator{\vol}{vol}
\DeclareMathOperator{\sign}{sign}

\newtheorem{problem}{Problem}
\begin{document}

\title%[Lagrangian transport through surfaces in volume-preserving flows]
{Lagrangian transport through surfaces in volume-preserving flows}
\date{\today}
\author{Daniel Karrasch\thanks{Current address: Technische Universität München, Zentrum Mathematik -- M3, Boltzmannstraße 3, 81937 Garching bei München, Germany} \footnote{Electronic mail: \href{mailto:karrasch@ma.tum.de}{karrasch@ma.tum.de}}\\ETH Zürich, Institute for Mechanical Systems\\ Leonhardstrasse 21, 8092 Zürich, Switzerland.}

%\pacs{05.45.-a, 05.60.Cd, 47.10.-g, 47.11.-j, 47.52.+j}

\maketitle

\begin{abstract}
Advective transport of scalar quantities through surfaces is of fundamental
importance in many scientific applications. From the Eulerian perspective
of the surface it can be quantified by the well-known integral of
the flux density. The recent development of highly accurate semi-Lagrangian
methods for solving scalar conservation laws and of Lagrangian approaches
to coherent structures in turbulent (geophysical) fluid flows necessitate
a new approach to transport from the (Lagrangian) material perspective.
We present a Lagrangian framework for calculating transport of conserved
quantities through a given surface in $n$-dimensional, fully aperiodic,
volume-preserving flows. Our approach does not involve any dynamical
assumptions on the surface or its boundary.
\end{abstract}

\section{Introduction}

The transfer of a quantity along the motion of some carrying fluid,
or \emph{Lagrangian transport} for short, is of fundamental importance
to a broad variety of scientific fields and applications. The latter
include (geophysical) fluid dynamics \cite{Ottino1989,Wiggins1992,Samelson2006,Lin2013},
chemical kinetics \cite{Deen2013}, fluid engineering \cite{Pouransari2010,Speetjens2014}
and plasma confinement \cite{Boozer2005}. Existing methods for computing
transport (i) have mostly been developed under certain assumptions on
temporal behavior (steady or periodic time-dependence) \cite{Mackay1984,MacKay1990,MacKay1994,RomKedar1990};
spatial location (regions related to invariant manifolds such as lobe dynamics and
dividing surfaces in transition state theory) \cite{Mackay1984,MacKay1990,MacKay1994,RomKedar1990,Malhotra1998,Balasuriya2006};
state space dimension (2D) \cite{Mackay1984,RomKedar1990,Malhotra1998,Balasuriya2006};
or (ii) restrict to a perturbation setting \cite{Balasuriya2006}.
Recently, the problem of quantifying finite-time transport in aperiodic
flows between distinct, arbitrary flow regions has been considered
by Mosovsky \emph{et al.} \cite{Mosovsky2013}. They present a framework for
describing and computing finite-time transport in $n$-dimensional (chaotic) volume-preserving
flows, which relies on the reduced dynamics of an ($n-2$)-dimensional
`minimal set' of fundamental trajectories. In this paper, we
present a Lagrangian approach to the complementary problem of computing transport
through a codimension-one surface over a finite-time interval
in volume-preserving flows. This cannot be reformulated in the setting
of \cite{Mosovsky2013}, since (i) initial and final positions of
surface-crossing particles are \emph{a priori} unknown, and (ii) particles
may cross the surface several times, possibly in opposite directions,
leading to a net number of surface crossings different from one, in
particular including zero.

The problem of computing the flux through a surface in general flows
admits a well-known solution in terms of an integral of the flux density
over the surface and the time interval of interest, cf.\ the left-hand
side of Eq.\ \eqref{eq:aim_of_game}. We view this approach as \emph{Eulerian},
in that it involves instantaneous information (flux density) at fixed
locations in spacetime. Recently, two lines of research emerged
that inevitably require a \emph{Lagrangian} approach to the flux calculation.

The first is concerned with the numerical solution of advection equations
(or, in the absence of sources, conservation laws) for conserved quantities
by semi-Lagrangian methods, which enjoy geometric flexibility and
the absence of Eulerian stability constraints \cite{Zhang2013a,Zhang2013b}.
Roughly speaking, the term semi-Lagrangian refers to methods which evolve
material densities on spatially fixed test volumes based on short-term
material advection steps.% In contrast, a fully Lagrangian approach
%would track the locations of material particles over extended times, and a fully Eulerian
%approach would monitor the evolution of densities as functions of spatial positions.

The second is concerned with transport by coherent vortices
in oceanic flows, more precisely with the determination of the relevance
of \emph{coherent transport} \cite{Dong2014,Zhang2014,Huhn2015},
an open problem in physical oceanography and climate science. Coherent
structures have long been studied in fluid dynamics \cite{FazleHussain1986},
typically from an Eulerian point of view, i.e., by defining coherent structures as 
subsets of \emph{spacetime}, usually as (sub-)level sets of scalar fields on spacetime 
\cite{McWilliams1984,Provenzale1999,Dong2014,Zhang2014}.
This approach, however, yields coherent structures with a priori unclear 
relation to actual fluid motion and hence
ambiguous role in coherent transport; see transfer-operator related approaches
which seek Eulerian coherent sets with minimal flux through the boundary under advection with 
small-scale diffusion superimposed, cf., for instance, \cite{Froyland2013}.
For these reasons, Lagrangian approaches to coherent structures have been developed
over the last few years. These seek coherent structures as subsets of the \emph{material} evolving 
under the flow \cite{Haller2015,Haller2013a,Karrasch2015,Hadjighasem2015,Froyland2015a,Froyland2015,Ma2014,Mundel2014},
building on a wide variety of mathematical principles.
The location of a material structure in spacetime is fully determined by its motion.

\begin{figure}
%preprint version
\centerline{\includegraphics[width=0.45\columnwidth]{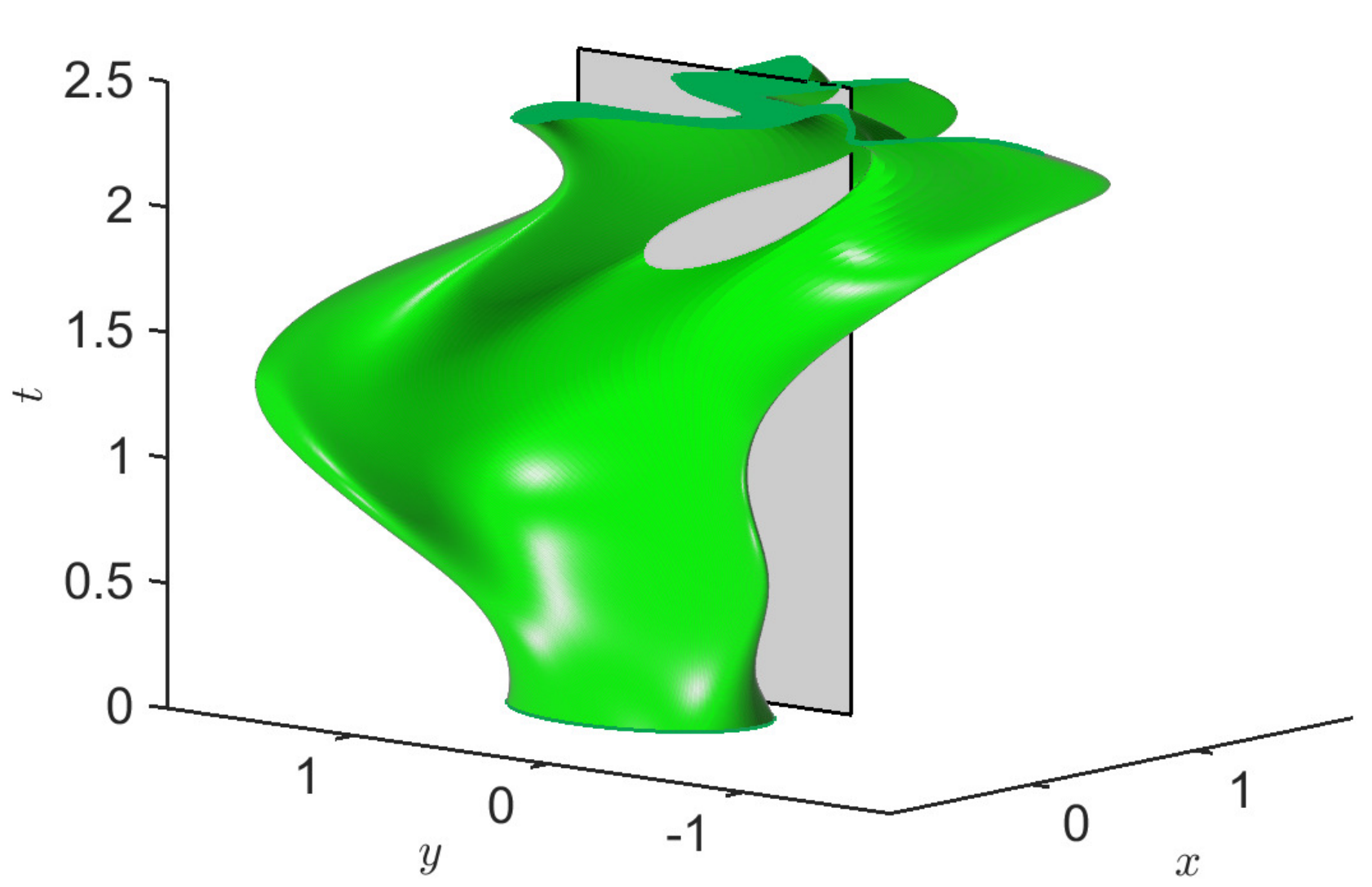}
\quad\includegraphics[width=0.45\columnwidth]{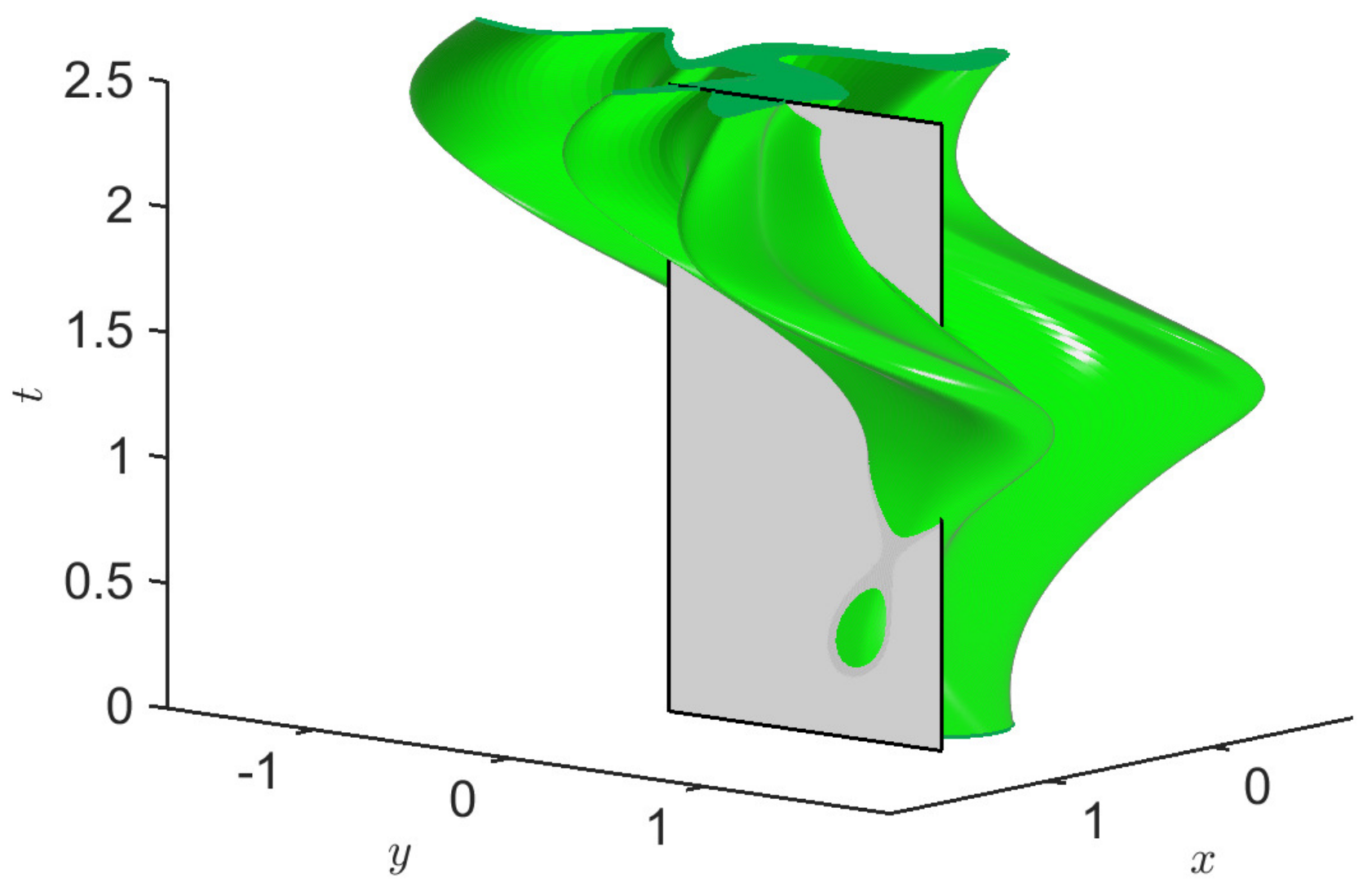}}
%reprint version
%\centering{\includegraphics[width=0.6\columnwidth]{flow_front.png}\\
%\includegraphics[width=0.6\columnwidth]{flow_back.png}}
\caption{Views from opposite directions on the evolution of an initial ellipse over time,
and its intersection with a surface in extended state space. Computing the flux
contribution by the material points originating from that ellipse in an Eulerian
framework corresponds to computing the flux integral over the subset of the section
which is contained in the interior of the cylindrical structure.}
\label{fig:flow}
\end{figure}

This implicit spatial definition of material structures poses a severe challenge to determining
their contribution to transport through a surface in an Eulerian manner. First, 
for each point on the extended surface in spacetime knowledge is required 
whether it is occupied by a particle originating from the material structure of interest. 
Second, the subset of points on the surface which are occupied by material 
particles of interest may be very complicated, cf.\ Fig.\ \ref{fig:flow},
especially in turbulent flows.
As a consequence, an Eulerian integration of the flux density restricted
to the intersection of the extended surface with the path of the material set
of interest is practically infeasible.

Conversely, from the material point of view, it is intuitive that
the number of surface crossings of each individual Lagrangian particle
is relevant to determine its flux contribution.
First steps towards a classification of Lagrangian particles with respect to
net number of curve crossings have been proposed for two-dimensional
flows under additional technical assumptions by Zhang \cite{Zhang2013,Zhang2015}. In
this paper, we solve the following more general \emph{donating region
problem} \cite[Def.~1.2]{Zhang2013b}.

%\subsubsection*{Problem statement}
\begin{problem}\label{problem}
For a given regular, divergence-free and time-dependent vector field
$\boldsymbol{u}(t,\boldsymbol{x})=\boldsymbol{u}_{t}(x)$, consider a conserved
quantity $f(t,x)=f_{t}(x)$, which satisfies the scalar conservation
law in Eulerian form %
\footnote{The Lagrangian form is $\frac{Df}{Dt}=0$ with $\frac{D}{Dt}$ the
material derivative. Physically, this means that the scalar $f$ does
not change along particle motions.%
}
\[
\partial_{t}f+\boldsymbol{u}_{t}\boldsymbol{\cdot\nabla} f=0.%\label{eq:conservation}
\]
Let $\mathcal{C}$ be a compact, connected, embedded codimension-one
surface in (configuration space) $\mathbb{R}^{n}$, and $\mathcal{T}=[0,\tau]$
be a compact time interval. The problem is to find pairwise disjoint
sets $\mathcal{D}_{k}\subset\mathbb{R}^{n}$, indexed by $k\in\mathbb{Z}$,
of Lagrangian particles at time $t=0$, such that
\begin{equation}
\int_{\mathcal{C}\times\mathcal{T}}f_t(x)\cdot\boldsymbol{u}_t(x)\boldsymbol{\cdot}\boldsymbol{n}
(x)\,\mathrm{d}x\,\mathrm{d}t=\sum_{k\in\mathbb{Z}}\thinspace k\cdot\int_{\mathcal{D}_{k}} f_{t}\bigr|_{t=0}(p)\,\mathrm{d}p,\label{eq:aim_of_game}
\end{equation}
where $\boldsymbol{n}$ is the unit normal vector field to $\mathcal{C}$
characterizing the direction of positive flux.
\end{problem}

In fact, we may generalize Problem \ref{problem} and drop the assumption that $\mathcal{C}$ 
be stationary over $\mathcal{T}$. Thus, we solve
the analogous problem for smoothly moving surfaces $\mathcal{H}=\bigcup_{t\in\mathcal{T}} \mathcal{C}_t$,
i.e., $\mathcal{H}$ is everywhere transversal to time fibers $\lbrace t\rbrace\times\mathbb{R}^n$.

\begin{problem}\label{problem2}
Let 
$\boldsymbol{u}(t,\boldsymbol{x})=\boldsymbol{u}_{t}(x)$ and $f(t,x)=f_{t}(x)$ be as in Problem \ref{problem}, and
$\hat{\boldsymbol{u}}_t(x)=\left(\begin{smallmatrix} 1\\ \boldsymbol{u}_t(x)\end{smallmatrix}\right)$ the
extended velocity field. Let $\mathcal{H}$ be a codimension-one
surface in extended configuration space $\mathcal{T}\times\mathbb{R}^{n}$, $\mathcal{T}=[0,\tau]$,
as defined above. The problem is to find pairwise disjoint
sets $\mathcal{D}_{k}\subset\mathbb{R}^{n}$, indexed by $k\in\mathbb{Z}$,
of Lagrangian particles at time $t=0$, such that
\begin{equation}
\int_{\mathcal{H}}f_t(x)\cdot\hat{\boldsymbol{u}}_t(x)\boldsymbol{\cdot}\hat{\boldsymbol{n}}_t
(x)\,\mathrm{d}(t,x)=\sum_{k\in\mathbb{Z}}\thinspace k\cdot\int_{\mathcal{D}_{k}} f_{t}\bigr|_{t=0}(p)\,\mathrm{d}p,\label{eq:aim_of_game2}
\end{equation}
where $\hat{\boldsymbol{n}}_t(x)$ is the unit normal vector field to $\mathcal{H}$ at $(t,x)$.
\end{problem}

%We will provide a short proof of the folklore identity
%\begin{equation}\label{eq:flux}
%\int_{\mathcal{S}}f_t(x)\cdot\hat{\boldsymbol{u}}_t(x)\boldsymbol{\cdot}\hat{\boldsymbol{n}}_t
%(x)\,\mathrm{d}(t,x) = \int_{\mathcal{T}}\int_{\mathcal{S}_t}f_t(x)\cdot(\boldsymbol{u}_t(x)-\boldsymbol{v}_t(x))\boldsymbol{\cdot}\boldsymbol{n}_t
%(x)\,\mathrm{d}x\,\mathrm{d}t,
%\end{equation}
%where $\mathcal{S}_t$ is the time-$t$ slice of $\mathcal{S}$, $\boldsymbol{v}_t(x)$ is the 
%virtual velocity of $\mathcal{S}$,\footnote{More precisely, $\boldsymbol{v}_t(x)$ is the spatial component
%of the tangent vector $\hat{\boldsymbol{v}}_t(x)$ of $\mathcal{S}$ at $(t,x)$, where $\hat{\boldsymbol{v}}_t(x)$
%is normal to $\mathcal{S}_t$ at $x$ and has time-component equal to 1.}
%and $\boldsymbol{n}_t(x)$ is the purely spatial 
%unit normal vector field to $\mathcal{S}_t$ at $x$, cf., \cite[Thm.\ 2.1 \& Fig.\ 2]{Froyland2015c}.

For ease of presentation, we will first focus on the stationary case as in Problem \ref{problem}, before we
discuss the general case as in Problem \ref{problem2} in Section \ref{sec:moving_surface}. 

\section{Illustrative discussion of the main result}\label{sec:discussion}

The aim of this section is to discuss  Eq.\ \eqref{eq:aim_of_game} in anticipation
of mathematical constructions presented in Sections Section \ref{sec:coordinates} and Section \ref{sec:Transport}.
In fact, those constructions emerge naturally when taking a differential
topological view on the change from Eulerian to Lagrangian coordinates. We recall
and discuss concepts related to that coordinate change in Section \ref{sec:coordinates}.% With this discussion at hand,
%the reader may skip Sections Section \ref{sec:coordinates} and Section \ref{sec:Transport} in
%a first reading, and proceed directly to the illustrating example in Section
%\ref{}

First, the left-hand side of  Eq.\ \eqref{eq:aim_of_game} corresponds to the
classical flux integral over the surface $\mathcal{C}$ over time $\mathcal{T}$.
The integrand, also called the flux density, corresponds to the normal component
of the velocity field weighted with the material density $f$.
Here, the flux is viewed from the perspective of the section, which is why we
refer to the left-hand side of  Eq.\ \eqref{eq:aim_of_game} as the Eulerian flux integral.

In contrast, the right-hand side of  Eq.\ \eqref{eq:aim_of_game} is an integral, or
a sum of integrals, solely over material points or particles.
As our later analysis reveals we may decompose the whole material domain (up to a measure zero set)
into disjoint material subsets, each of which contains particles with a certain
number of net crossings across the surface $\mathcal{C}$ over the time interval
$\mathcal{T}$. These material subsets are denoted by $\mathcal{D}_k$, where $k$
refers to the net number of transversal crossings, and have been coined \emph{donating regions
of fluxing index $k$} by Zhang \cite{Zhang2015}. The contribution to transport
by each donating region corresponds to its respective mass (or volume in a homogeneous fluid) multiplied by
its corresponding fluxing index. The overall transport through $\mathcal{C}$
is then given by the sum of transport contributions of all donating regions. In summary,
the right-hand side of  Eq.\ \eqref{eq:aim_of_game} views the flux from the perspective
of the mass-carrying particles, which is why we refer to it as the Lagrangian
flux integral.

For illustration, consider an array of vortices, given by the stream function
\[
H(x,y)=-A\sin\left(\pi x\right)\sin\left(\pi y\right),
\]
subject to an aperiodic, spatially uniform spiraling forcing given by 
\[
F(t)=\left(t\sin\left(\pi t\right), t\cos\left(\pi t\right)\right)^{\top}.
\]
The velocity field is hence
\[
\boldsymbol{u}(t,x,y)=\begin{pmatrix}\phantom{-}\partial_{y}H(x,y)-t\sin\left(\pi t\right)\\ -\partial_{x}H(x,y)-t\cos\left(\pi t\right)\end{pmatrix}.%^{\top}.
\]

\begin{figure}
\centerline{\includegraphics[height=.32\textheight]{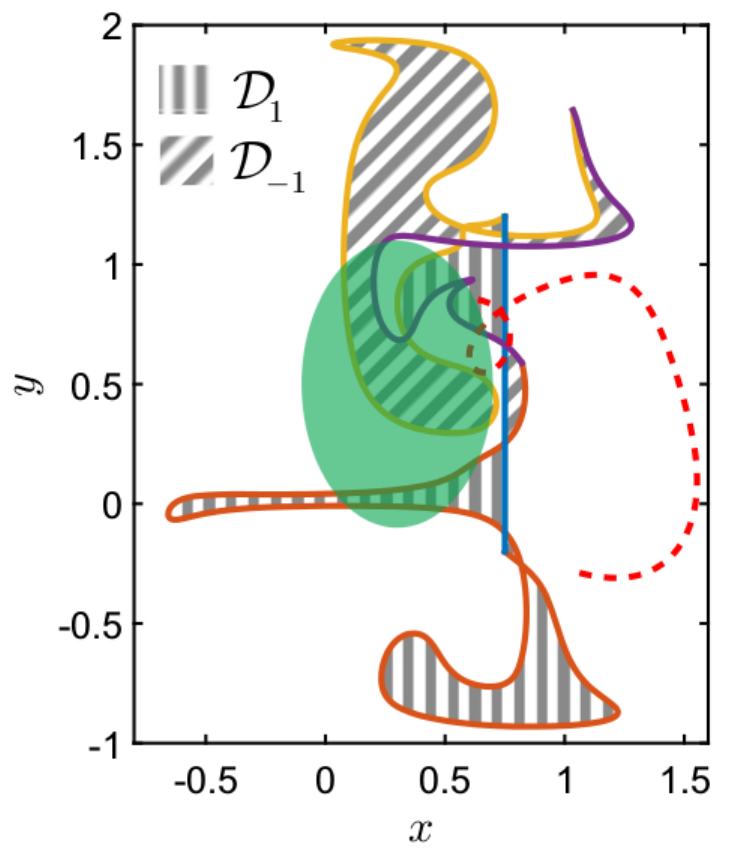}}
\caption{Donating regions divided in cells with fluxing index
$+1$ (vertically hatched), $-1$ (diagonally hatched) and $0$ (white).
The dashed curve (red) is a sample trajectory (path line).
%For later reference, the section $\mathcal{C}$ (blue), the streak lines through the boundary
%points of $\mathcal{C}$ (yellow \& brown), and the backward image
%of the section (purple) correspond to the curves shown of same color in Figs.\ Fig.\ \ref{fig:Euler}
%and Fig.\ \ref{fig:Lagrange}.
The ellipse (dark green) corresponds to the
set of Lagrangian particles whose flow evolution is shown in Fig.\ \ref{fig:flow}.}
\label{fig:donating_region}
\end{figure}

We set the section to
$\mathcal{C}=\left\{ x=0.75,\thinspace-0.2\geq y\geq1.2\right\}$,
the time interval to $\mathcal{T}=\left[0,2.5\right]$, and, for simplicity, the material density $f\equiv 1$.
Fig.\ \ref{fig:donating_region}
shows the regions of particles with fluxing index $+1$, $0$, and $-1$; here,
white corresponds to fluxing index $0$. For details on how to construct
the donating regions, see  Section \ref{sec:2d}. There
are no particles with other fluxing indices, and some particles attain a
fluxing index $+1$ by crossing twice in positive and once in negative direction,
such as the sample trajectory shown in red.

For a numerical validation of  Eq.\ \eqref{eq:aim_of_game}, we compute its
left-hand side by numerical quadrature, which
yields a transport value of $-0.142$. For the right-hand side, we
subtract the area of $\mathcal{D}_{-1}$ from that of $\mathcal{D}_1$, and
obtain a transport value of $-0.141$, with relative error below $1\%$. For
further improvements of area calculations via spline-approximation of
polygon boundaries see \cite{Zhang2013a}. Note that, in practice, the
numerical integration of the Eulerian integral becomes more challenging
in case the section $\mathcal{C}$ is curved and requires a parametrization.
In contrast, we will see that a curved section does not add any difficulty
in the Lagrangian framework. This geometric flexibility has been one major motivation to develop
semi-Lagrangian numerical schemes for solving conservation laws, cf.\ \cite{Zhang2013a,Zhang2015}.

For our main intended application, the calculation of the contribution to transport
by a certain material subset $A$---such as the ellipse in
Figs.\ \ref{fig:flow} and \ref{fig:donating_region} (dark green, centered at $(0.3,0.5)$ with semi-major axis $0.6$ in $y$-direction, 
and semi-minor axis $0.4$ in $x$-direction)---reduces to subtracting the area of $A\cap\mathcal{D}_{-1}$ from $A\cap\mathcal{D}_{1}$,
which yields $-0.1061$.

At first sight, this may seem paradoxical, since the entire ellipse is
launched from one side of the section. Fig.\ \ref{fig:flow} now indicates
that at the end of the time interval, parts of the material ended up on the
other side of the section. Going from the first to the latter side
corresponds to positive flux, and one would expect an overall positive
transport contribution. However, a closer inspection of Fig.\ \ref{fig:flow}
reveals that most of the material which is advected to the other side has actually not crossed
the section, but has flown around it. Moreover, much of the material which
has flown around has actually crossed the section in negative direction, thus
yielding an overall negative flux contribution of the ellipse.

The rest of this paper is devoted to the derivation of 
Eqs.\ \eqref{eq:aim_of_game} and \eqref{eq:aim_of_game2}.
Most of the involved abstract concepts turn out to have a
nice physical interpretation.
%In Lagrangian terms, restriction to a given set of Lagrangian particles $A$
%of interest is then conveniently performed by intersection of the
%$\mathcal{D}_k$'s with $A$, see also Fig.\ Fig.\ \ref{fig:donating_region}.

\section{Eulerian and Lagrangian coordinates}\label{sec:coordinates}

While Eulerian coordinates $x$ are assigned to \emph{spatial} points
in a fixed frame of reference, Lagrangian coordinates $p$ label \emph{material}
points and are usually taken as the Eulerian coordinates at some initial
time, say, $t=0$. The motion of material points is described by the
flow $\varphi$, a mapping between initial positions $p$ at time $t=0$
and current positions $x$ at time $t$, i.e., $\varphi_{0}^{t}(p)=x$.
Thus, the flow map $\varphi_{0}^{t}$ can be interpreted as a change
from Lagrangian to Eulerian coordinates. The inverse flow
map $\varphi_{t}^{0}=\left(\varphi_{0}^{t}\right)^{-1}$ corresponds then to the
change from Eulerian to Lagrangian coordinates.

In fluid dynamics, there are two important characteristic curves associated
with the flow \cite{Batchelor2000}, which we re-interpret in terms
of Eulerian and Lagrangian coordinates.

A \emph{path line through $p$} is the time-curve of a fixed Lagrangian
particle $p$ in Eulerian coordinates, i.e., $t\mapsto\varphi_{0}^{t}(p)$.
In other words, the path line is a collection of Eulerian positions
that the Lagrangian particle $p$ will occupy at some time. Its time
derivative, expressed in Eulerian coordinates, gives rise to the \emph{velocity
field}
\begin{equation}
\boldsymbol{u}_{t}(x)\coloneqq\boldsymbol{u}(t,x)=\partial_{t}\varphi_{0}^{t}(\varphi_{t}^{0}(x)).\label{eq:ODE}
\end{equation}
By construction, the path line through $p$ is the solution of the
initial value problem
\begin{align*}
\dot{x} & =\boldsymbol{u}_{t}(x), & x(0) & =p.
\end{align*}

\begin{figure}
\centerline{\includegraphics[height=0.25\textheight]{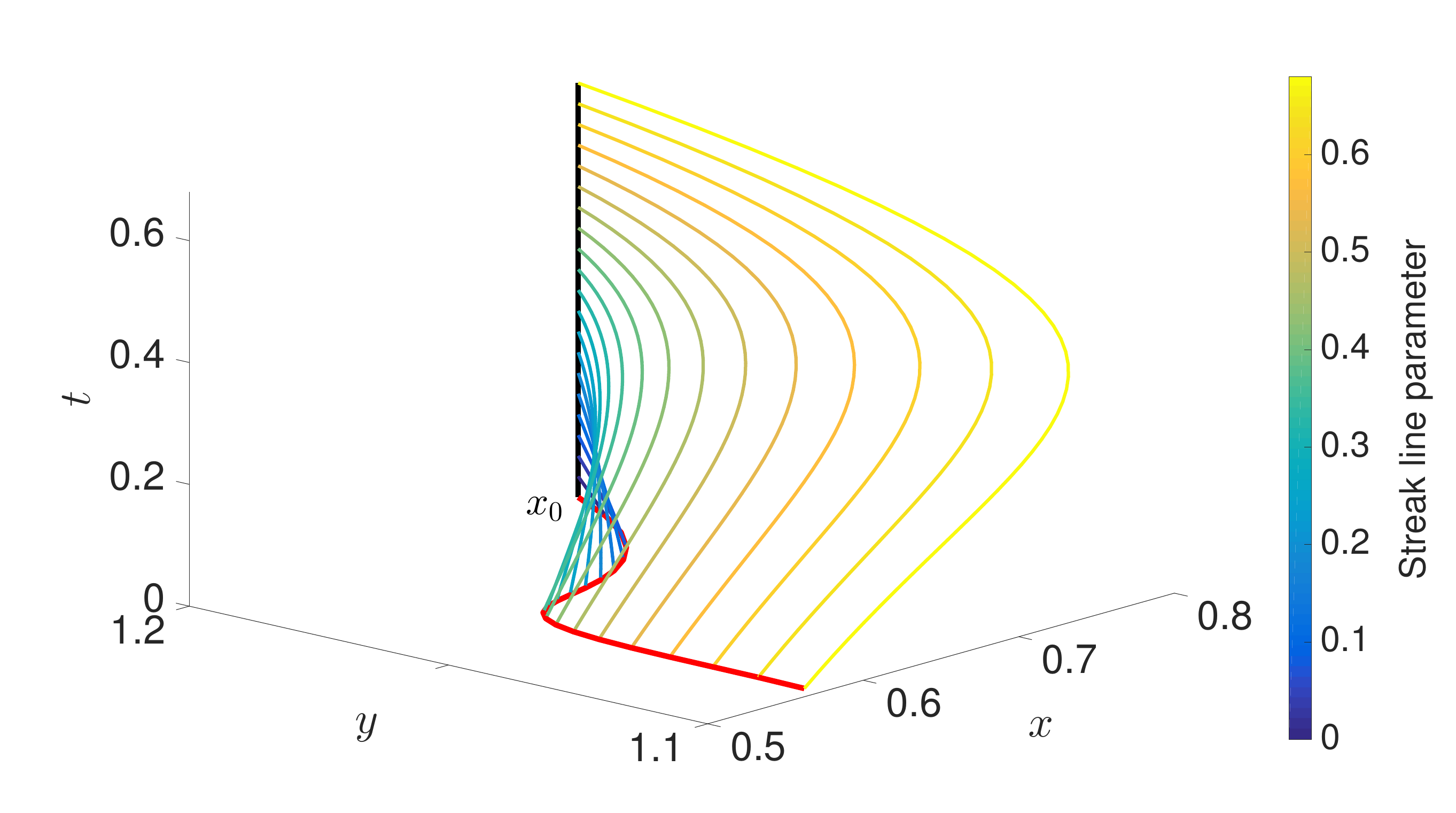}}
%reprint version
%\centering{\includegraphics[width=0.6\columnwidth]{flow_front.png}\\
%\includegraphics[width=0.6\columnwidth]{flow_back.png}}
\caption{Streak line and streak vector field: the streak line through $x_0$ (red) is constructed by 
subsequently advecting $x_0$ from time $t$ backwards to time 0. Conversely, when a
particle from the streak line, with curve parameter
$s$ (colorbar), is launched at time 0 and advected $s$ time units, the particle reaches 
the spatial position $x_0$ (black). The streak vector field corresponds to the velocity of the streak lines.
The bottom $x$-$y$-plane is to be considered as Lagrangian coordinates. Correspondingly,
the streak vector field is a material vector field, in contrast to the classic velocity field, which
is a spatial vector field.}
\label{fig:streak_vf}
\end{figure}

A \emph{streak line through $x$} is the time-curve of a fixed Eulerian
location $x$ in Lagrangian coordinates, i.e., $t\mapsto\varphi_{t}^{0}(x)$.
In other words, the streak line is a collection of material points
that will occupy the Eulerian position $x$ at some time, see Fig.\ \ref{fig:streak_vf}. Our definition of streak lines
suits well the currently intended purpose and is consistent with Zhang's use \cite{Zhang2015}.
It is more common and actually more intuitive, however,
to view a streak line as the collection of material points that have passed
the Eulerian position $x$ at some time. In this context, a streak line can be 
imagined as an instantaneous curve of Lagrangian markers, injected in the past at $x$
and passively advected by the flow, see \cite{Batchelor2000}. The time
derivative of the streak line, expressed in Lagrangian coordinates, gives rise to the
\emph{streak vector field}
\[
\boldsymbol{w}_{t}(p)\coloneqq\boldsymbol{w}(t,p)=\partial_{t}\varphi_{t}^{0}(\varphi_{0}^{t}(p)).
\]
By construction, the streak line through $x$ is the solution of the
initial value problem
\begin{align*}
\dot{p} & =\boldsymbol{w}_{t}(p), & p(0) & =x.
\end{align*}

To derive an algebraic relation between the velocity
field $\boldsymbol{u}$ and the streak vector field $\boldsymbol{w}$, we
compute the velocity $\partial_{t}\varphi_{t}^{0}(x)$ along the streak line 
$t\mapsto\varphi_{t}^{0}(x)$ through $x$. Differentiation
of the constant function $t\mapsto\varphi_{0}^{t}\left(\varphi_{t}^{0}(x)\right)=x$
with the shorthand notation $p=\varphi_{t}^{0}(x)$ yields
\[
0=\partial_{t}\varphi_{0}^{t}\left(p\right)+\mathrm{d}\varphi_{0}^{t}\left(p\right)\partial_{t}\varphi_{t}^{0}\left(x\right).
\]
Using Eq.\ \eqref{eq:ODE} and the invertibility of the linearized flow map with 
$\mathrm{d}\varphi_{0}^{t}\left(p\right)^{-1}=\mathrm{d}\varphi_{t}^{0}\left(x\right)$,
we obtain
\begin{equation}
\begin{split}\boldsymbol{w}_{t}(p)= & -\left(\mathrm{d}\varphi_{0}^{t}\left(p\right)\right){}^{-1}\boldsymbol{u}\left(t,\varphi_{0}^{t}\left(p\right)\right).\end{split}
\label{eq:streak_vf}
\end{equation}
Equivalently, in global terms we have
\begin{align*}
\boldsymbol{w}_{t} & =-\left(\varphi_{t}^{0}\right)_{*}\boldsymbol{u}_{t}, & \boldsymbol{u}_{t} & =-\left(\varphi_{0}^{t}\right)_{*}\boldsymbol{w}_{t},
\end{align*}
where the subindex $*$ denotes pushforward of vector fields by $\varphi_{t}^{0}$
and $\varphi_{0}^{t}$, respectively. Eq.\ \eqref{eq:streak_vf} is an alternative to the formula originally derived
by Weinkauf \& Theisel \cite{Weinkauf2010}.

Finally, we recall that the flow $\varphi$ is volume-preserving on $\mathbb{R}^{n}$
if and only if the velocity field $\boldsymbol{u}$ (or, equivalently,
$\boldsymbol{w}$) is divergence-free \cite{Batchelor2000}.

\section{Transport through surfaces}\label{sec:Transport}

In this section, we solve Problem \ref{problem}
by analyzing the change from Eulerian to La\-grang\-ian coordinates
from the differential topology viewpoint \cite{Milnor1965,Hirsch1976}.

\subsection{Setting}

\begin{figure}
\centerline{\includegraphics[height=.28\textheight]{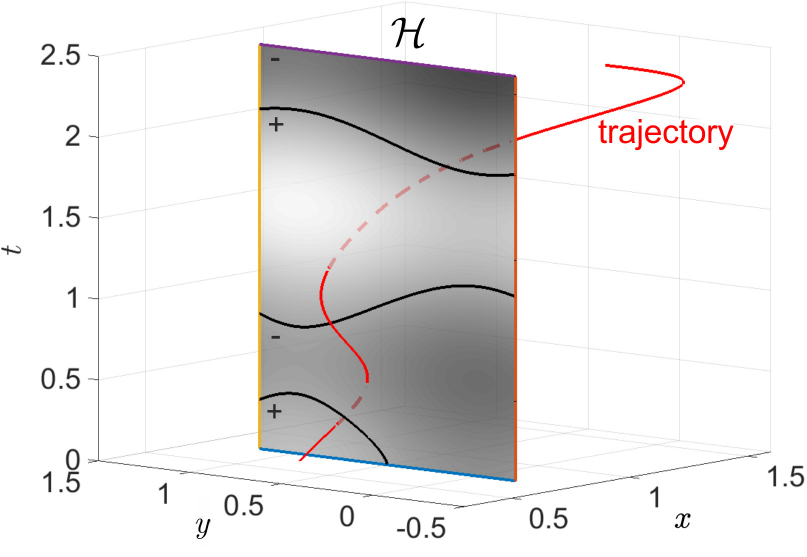}}
\caption{Eulerian coordinates: Flux density on the extended section
$\mathcal{H}$, bounded by blue, yellow, purple and brown lines. Positive
flux is directed in positive $x$-direction, and the flux direction through
$\mathcal{H}$ is indicated by $+/-$. Zero-level curves of flux are shown in black.
Also shown is the extended pathline from Fig.\ \ref{fig:donating_region} (red, dashed when behind the
section) with three transversal crossings, two positive and one negative.}
\label{fig:Euler}
\end{figure}

\begin{figure}
\centerline{\includegraphics[height=.28\textheight]{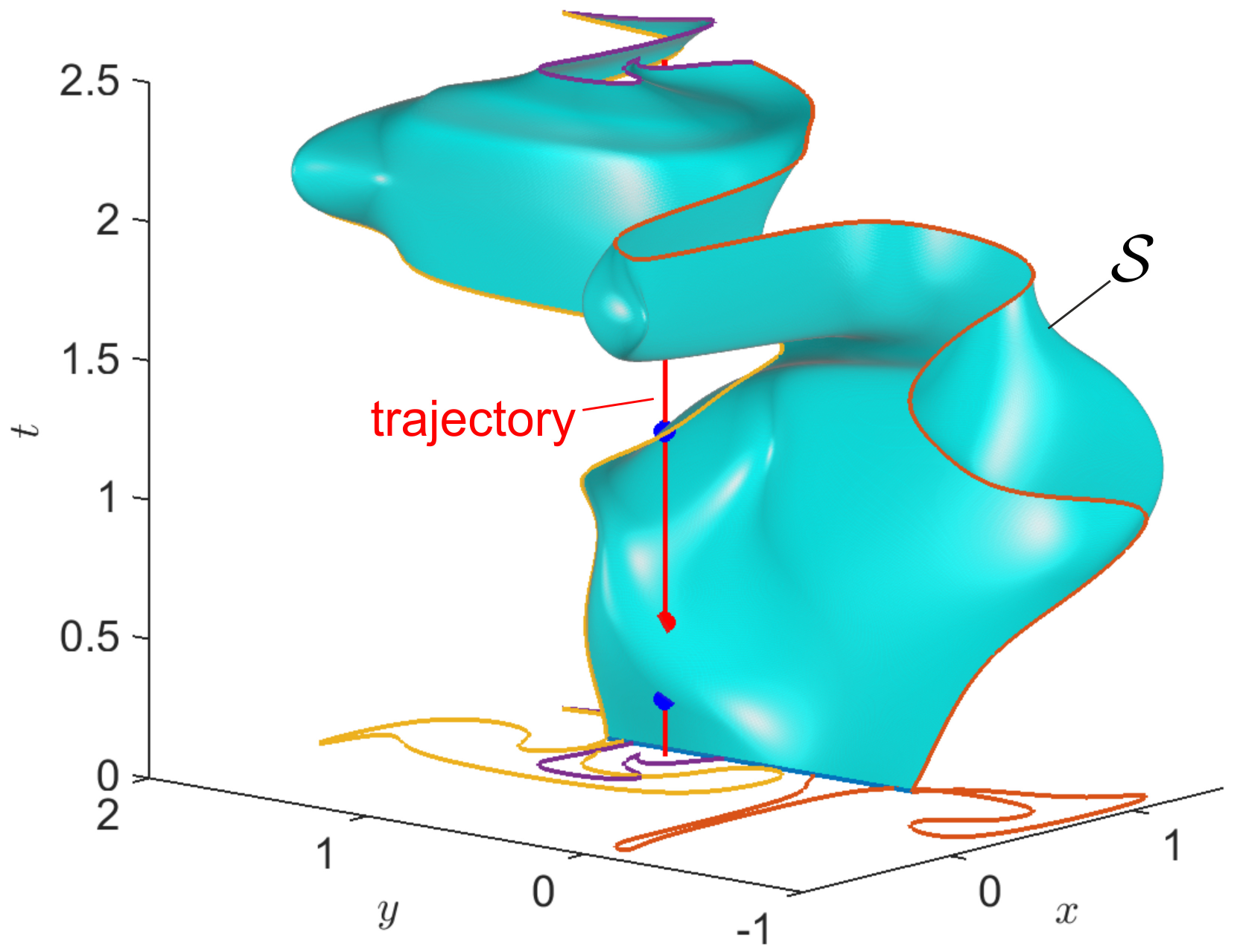}}
\caption{Lagrangian coordinates: Streak surface $\mathcal{S}$ and trajectory
(red), which crosses $\mathcal{S}$ twice in positive
(blue balls) and once in negative (red ball) direction. The boundary
curves correspond to the boundary curves in Fig.\ \ref{fig:Euler}
with the respective colors. Their projections onto the $t=0$ time
slice are shown in the corresponding colors, cf., Fig.\ \ref{fig:donating_region}.}
\label{fig:Lagrange}
\end{figure}

We decompose the change from Eulerian to Lagrangian coordinates (evaluated 
along the extended section
$\mathcal{T}\times\mathcal{C}$) into two steps. First, we map Eulerian spacetime
points to their respective
initial position at time $t=0$, while keeping them on the same time
slice, i.e.,
\begin{align*}
\Phi\colon\mathcal{T}\times\mathbb{R}^{n} & \to\mathcal{T}\times\mathbb{R}^{n}, & \left(t,x\right) & \mapsto\left(t,\varphi_{t}^{0}(x)\right)=(t,p).
\end{align*}
The transformation $\Phi$ maps the (extended) section $\mathcal{H}\coloneqq\mathcal{T}\times\mathcal{C}$,
see Fig.\ \ref{fig:Euler}, diffeomorphically to the \emph{streak surface} $\mathcal{S}\coloneqq\Phi(\mathcal{H})$,
see Fig.\ \ref{fig:Lagrange}.

In extended Eulerian coordinates, trajectories (or, extended path
lines) take the form $(t,\varphi_{0}^{t}(p))$; in extended Lagrangian
coordinates, trajectories are simply vertical lines, i.e., $t\mapsto(t,p)$,
since they track the same particle $p$,
see Fig.\ \ref{fig:Lagrange}. Intersections of path lines with $\mathcal{H}$
correspond one-to-one with intersections of $\mathcal{S}$ with vertical
lines.

In a second step, we project spacetime points to their spatial coordinates
by the canonical projection $\Pi\colon\mathcal{T}\times\mathbb{R}^{n}\to\mathbb{R}^{n}$
and introduce $\mathcal{D}\coloneqq\Pi(\mathcal{S})$. We view the
image of $\Pi$ as the initial time slice of the extended state space
and therefore parametrized by Lagrangian coordinates. By construction,
we have for $\Psi\coloneqq\Pi\circ\Phi$ the identity $\Psi(t,x)=\varphi_{t}^{0}(x)$,
and it is exactly the particles from $\mathcal{D}$ that cross $\mathcal{C}$
within $\mathcal{T}$ one way or another, possibly multiple times.

For later reference, we emphasize the following observation. Let $\boldsymbol{e}_t$
be the unit time-like vector field. Then we have
\begin{subequations}\label{eq:action}
\begin{align}
\mathrm{d}\phi_0^t\mathrm{d}\Psi(t,x)\boldsymbol{e}_t(x) &= -\boldsymbol{u}_t(x).\label{eq:action1}
\intertext{On the other hand, for any space-like vector field $\boldsymbol{v}_t$ such as, for instance,
the velocity field $\boldsymbol{u}$, we have}
\mathrm{d}\phi_0^t\mathrm{d}\Psi(t,x)\boldsymbol{v}_t(x) &= \boldsymbol{v}_t(x),\label{eq:action2}
\end{align}
\end{subequations}
since in this case $\mathrm{d}\Psi$ acts like $\mathrm{d}\phi_t^0$.

\subsection{Counting net crossings -- the degree}

We study the differentiable map $\Psi$ as a map between the two $n$-dimensional
manifolds $\mathcal{H}\subset\mathcal{T}\times\mathbb{R}^{n}$ and
$\mathcal{D}\subset\mathbb{R}^{n}$. To this end, we recall notions
from differential topology and interpret them in our setting.

First, \emph{regular points} $\left(t,x\right)\in\mathcal{H}$ are
those for which the differential $\mathrm{d}\Psi(t,x)$ is invertible.
This is exactly the case, when the tangent space at $\Phi(t,x)=(t,\phi_t^0(x))$ to $\mathcal{S}$
does not contain the vertical time direction. Non-regular points
$(t,x)$ of $\Psi$ are referred to as \emph{critical points}, which are
characterized by non-transversal crossings, i.e.,
%$\boldsymbol{u}_{t}(x)\boldsymbol{\cdot}\boldsymbol{n}(x)=0$,
instantaneous stagnation points $\boldsymbol{u}_{t}(x)=0$,
or tangential crossing/rebound.

According to Sard's theorem, the set of \emph{critical values}, i.e., images
of critical points, has measure zero in $\mathcal{D}$, even though
the set of critical points may be large in $\mathcal{H}$.
Consistently, critical points do not contribute to the Eulerian flux
integral.

In the Eulerian setting, positive/negative crossings of a Lagrangian
particle $p$ through $\mathcal{C}$ correspond to $\pm\boldsymbol{u}_t(x)\boldsymbol{\cdot}\boldsymbol{n}(x)>0$
at corresponding intersections $(t,x)$. In the Lagrangian setting, positive/negative
crossings are equivalent to crossings of the vertical line over $p$
through $\mathcal{S}$ from below/above with respect to the orientation on
$\mathcal{S}$, see Fig.\ \ref{fig:Lagrange}.
This is formalized by $\pm\det\mathrm{d}\Psi>0$, which geometrically
means that $\mathrm{d}\Psi(t,x)$ is orientation-preserving
or orientation-reversing at
the respective crossings of $p$. This is well-defined at all
crossings exactly for particles $p$ which are regular values of $\Psi$.% Due to the compactness of $\mathcal{H}$ the number
%of crossings of any regular value is finite.

The difference of positive and negative crossings, or, in other words, the
net number of crossings, is given by the \emph{degree of $\Psi$ at $p\in\mathcal{D}$},
\[
\deg\left(\Psi,p\right)\coloneqq\sum_{(t,x)\in\Psi^{-1}(p)\cap\mathcal{H}}\sign\left(\det\left(\mathrm{d}\Psi(t,x)\right)\right).
\]
The degree function is locally constant, and can be used to define
a partition of the set of particles, up to the measure-zero set of critical values. 
Specifically, we define $\mathcal{D}_{k}$
as the set of regular values $p$ of $\Psi$ such that $\deg\left(\Psi,p\right)=k$
for $k\in\mathbb{Z}$. Following \cite{Zhang2015}, we refer to the $\mathcal{D}_{k}$'s
as \emph{donating regions of fluxing index $k$}.
By construction, each $\mathcal{D}_{k}$ contains the particles
with only transversal crossings of which there are $k$ net. In particular,
all material points that do not cross $\mathcal{C}$ within $\mathcal{T}$
are contained in $\mathcal{D}_{0}$.

\subsection{Area formula}

In the homogeneous case $f\equiv1$, we have the following
identity for the degree of $\Psi$ \cite[Section 3.1.5, Thm.~6]{Giaquinta1998},
which is a consequence of the \emph{area formula} (\cite{Federer1996}
and \cite[Thm.~5.3.7]{Krantz2008}):
\begin{align*}
\int_{\mathcal{H}}\det\mathrm{d}\Psi(t,x)\,\mathrm{d}x\,\mathrm{d}t & =\int_{\mathbb{R}^{n}}\deg(\Psi,p)\,\mathrm{d}p=\sum_{k\in\mathbb{Z}}\thinspace k\vol(\mathcal{D}_{k}).
\end{align*}
The area formula is a generalization of the change of variables (in
integral) formula to non-injective maps such as $\Psi$. Clearly, the degree measures
the non-injectivity, and is equal to $\pm 1$ globally for orientation-preserving and -reversing
diffeomorphisms, respectively.

Inhomogeneous conserved quantities are con\-stant along trajectories in
volume-preserving flows. Therefore, we have $f_{t}(x)=f_{0}(\Psi(t,x))$,
and we conclude with the general area formula:
%preprint version
\[
\int_{\mathcal{H}}f_{t}(x)\det\mathrm{d}\Psi(t,x)\,\mathrm{d}x\,\mathrm{d}t=
\int_{\mathcal{H}}f_{0}(\Psi(t,x))\det\mathrm{d}\Psi(t,x)\,\mathrm{d}x\,\mathrm{d}t
=\sum_{k}\thinspace k\int_{\mathcal{D}_{k}}f_{0}(p)\,\mathrm{d}p.
\]
%reprint version
%\begin{multline*}
%\int_{\mathcal{H}}f_{t}(x)\det\mathrm{d}\Psi(t,x)\mathrm{d}t\mathrm{d}x=\\
%=\int_{\mathcal{H}}f_{0}(\Psi(t,x))\det\mathrm{d}\Psi(t,x)\mathrm{d}t\mathrm{d}x=\\
%=\sum_{k}\thinspace k\int_{\mathcal{D}_{k}}f_{0}(p)\mathrm{d}p,
%\end{multline*}
To prove Eq.\ \eqref{eq:aim_of_game} and thereby solve Problem \ref{problem},
it remains to show 
\begin{equation}
\det\mathrm{d}\Psi(t,x)=\boldsymbol{u}_{t}(x)\boldsymbol{\cdot}\boldsymbol{n}(x),\label{eq:detdPsi}
\end{equation}
which is done in Appendix Section \ref{sec:calculation}, taking advantage of the relation \eqref{eq:streak_vf}
between the streak vector field and the velocity field.

%\subsection{The curvilinear case}
%
%Our result extends to the case when the state space is an $n$-dimensional,
%oriented smooth manifold $\mathcal{M}$ with volume form $\omega$.
%All differential topological notions and arguments work naturally
%in the manifold setting.
%Note also that it does not require a (Riemannian) metric on $\mathcal{M}$.
%In the metric-free context, the Eulerian flux density is given by
%the interior product of $\boldsymbol{u}$ and $\omega$ (or $\tilde{\omega}$),
%often denoted by $i_{\boldsymbol{u}_{t}}\omega$ or $\boldsymbol{u}_{t}\thinspace\lrcorner\ \omega$.

\subsection{The two-dimensional case}\label{sec:2d}

Our results for volume-preserving flows on $\mathbb{R}^{n}$ simplify
in the two-dimensional case considered by Zhang \cite{Zhang2013,Zhang2015}
as follows. It is well-known that, in our context, the degree of
$\Psi$ at some regular value $p\in\mathcal{D}\subset\mathbb{R}^{2}$
equals the \emph{winding number} of $p$ with respect to the closed curve
$\Psi(\partial\mathcal{H})$ around $p$ \cite[Section 6.6]{Deimling1985}.
The winding number counts the net number of turns of
$\Psi(\partial\mathcal{H})$ around $p$ under one counter-clockwise
passage through $\partial\mathcal{H}$, see Fig.\ \ref{fig:donating_region}.

In the 2D case, we find the donating regions $\mathcal{D}_k$ of fluxing index $k$ as
follows. The boundary $\partial\mathcal{H}$ of the extended section $\mathcal{H}$
is a closed curve, just as its image under $\Psi$. In Figs.\ \ref{fig:Lagrange}
and \ref{fig:donating_region} its image corresponds to the concatenation of
the section (blue), one streakline
(say, yellow), the backward image of the section (purple) and the other
streakline (brown). This closed curve gives rise to possibly several connected
components, \emph{simple loops}, through its self-intersections. Now, the winding number is
constant on the interior of each simple loop, and it suffices to compute the winding number
for some contained sample point, which is a standard task in computational geometry.
As anticipated in Section \ref{sec:discussion},
in the case of constant density the Lagrangian flux calculation reduces
to the computation of the enclosed area of each component,
multiplication by its winding number and final summation over the loops.

In summary, we have generalized the main result of \cite{Zhang2015},
in which only the issue of counting net crossings was treated under
additional assumptions from a different methodological perspective.
Note also that for the isolated counting aspect, our characterization
in terms of $\deg\left(\Psi\right)$, or winding number in 2D, is also
valid in the case of a \emph{compressible} velocity field $\boldsymbol{u}$.

\section{Moving surfaces}\label{sec:moving_surface}

To find a Lagrangian flux integral in the moving surface case, Problem \ref{problem2},
we may proceed as in  Section \ref{sec:Transport} with $\mathcal{H}=\mathcal{T}\times\mathcal{C}$ replaced by 
$\mathcal{H}=\bigcup_t\mathcal{C}_t$ as the domain of the maps $\Phi$ and $\Psi$. 
All constructions and arguments work just as well as before. It
remains to find an expression analogous to Eq.\ \eqref{eq:detdPsi} for $\det\mathrm{d}\Psi(t,x)$:
\[
\det\mathrm{d}\Psi(t,x)=\hat{\boldsymbol{u}}_t(x)\boldsymbol{\cdot} \hat{\boldsymbol{n}}_t(x),
\]
which is shown in Appendix Section \ref{sec:calculation2}.
%\[
%\int_{\mathcal{H}}f_t(x)\cdot\hat{\boldsymbol{u}}_t(x)\boldsymbol{\cdot}\hat{\boldsymbol{n}}_t
%(x)\,\mathrm{d}(t,x)=\sum_{k\in\mathbb{Z}}\thinspace k\cdot\int_{\mathcal{D}_{k}} f_{t}\bigr|_{t=0}(p)\,\mathrm{d}p.
%\]

\section{Conclusion}

In this work we have devised a Lagrangian approach to transport through (co\-di\-men\-sion-one) surfaces in general $n$-dimensional, unsteady, volume-preserving flows. Studying the change from Lagrangian to Eulerian coordinates (and vice versa) alone, we have (i) discovered a striking analogy between path lines and streak lines, (ii) derived an algebraic relation between their associated velocity fields, (iii) found a natural way of determining the net number of surface crossings for individual Lagrangian particles up to a measure zero set, and, finally, (iv) transformed the Eulerian flux integral into a Lagrangian one. Thus, we have solved the donating region problem \cite{Zhang2013b} for volume-preserving flows in arbitrary finite dimension and smoothly moving surfaces. An obvious extension of the methodology is to handle non-volume-preserving flows, and work is in progress in this direction.

Besides the theoretical insights, the Lagrangian approach to transport through surfaces is of major importance in at least two relevant fluid mechanical applications. First, it provides the missing theoretical foundation of a family of highly accurate semi-Lagrangian finite volume and interface tracking methods \cite{Zhang2013a,Zhang2013b}. Second, it facilitates the efficient computation of transport by Lagrangian coherent vortices in large-scale oceanic flows \cite{Karrasch2015,Huhn2015}.

\appendix

\section{Proof of \texorpdfstring{$\det\mathrm{d}\Psi=\boldsymbol{u\cdot n}$}{det(dPsi)=u*n}}
\label{sec:calculation}

We show that $\det\mathrm{d}\Psi(t,x)=\boldsymbol{u}_{t}(x)\boldsymbol{\cdot}\boldsymbol{n}(x)$.
To this end, fix a regular point $(t,x)\in\mathcal{H}$. Then $\Psi$
acts diffeomorphically between an open neighborhood $\mathcal{U}$
of $(t,x)$ in $\mathcal{H}$ and its image $\mathcal{V}\coloneqq\Psi(\mathcal{U})$
in $\mathcal{D}$. We may choose local coordinates on $\mathcal{U}$
such that in $(t,x)$ we have an orthonormal basis, i.e., $\boldsymbol{e}_{t}\in T_{t}\mathcal{T}$,
$\linhull\left\{\boldsymbol{e}_{2},\ldots,\boldsymbol{e}_{n}\right\} =T_{x}\mathcal{C}$ and
$\boldsymbol{e}_{1}\in T_{x}^{\perp}\mathcal{H}$, with
\[
\bigwedge_{i=1}^n\boldsymbol{e}_{i}=\boldsymbol{e}_{1}\wedge\ldots\wedge\boldsymbol{e}_{n}, \qquad 
\boldsymbol{e}_{t}\wedge\boldsymbol{e}_{2}\wedge\ldots\wedge\boldsymbol{e}_{n}, \qquad \text{and}\qquad
\boldsymbol{e}_{t}\wedge\boldsymbol{e}_{1}\wedge\ldots\wedge\boldsymbol{e}_{n},
\]
unit volume parallelepipeds. To avoid a discussion on delicate orientation
issues, the following calculations involving determinants are to be
read up to sign. To avoid notational clutter, we use the $\wedge$-notation to denote both the spanned
parallelepipeds and their volume, computed as the determinant of the matrix with
columns given by the factors of the wedge product.

%Next, we introduce a basis of $T_{p}\mathcal{D}$ by pushing forward
%the $\boldsymbol{e}_{i}$'s, i.e., $\boldsymbol{f}_{i}\coloneqq\left(\varphi_{t}^{0}\right)_{*}\boldsymbol{e}_{i}$,
%which in turn span a unit space volume as well due to volume-preservation
%by $\varphi$. %:
%\begin{align*}
%\mathrm{d}p\left(\boldsymbol{f}_{1},\ldots,\boldsymbol{f}_{n}\right) & =\left(\varphi_{0}^{t}\right)^{*}\mathrm{d}x\left(\left(\varphi_{t}^{0}\right)_{*}\boldsymbol{e}_{1},\ldots,
%\left(\varphi_{t}^{0}\right)_{*}\boldsymbol{e}_{n}\right)\\
% & =\mathrm{d}x\left(\boldsymbol{e}_{1},\ldots,\boldsymbol{e}_{n}\right)=1.
%\end{align*}
%In the chosen coordinates, we get the following coordinate representation
%for $\boldsymbol{u}_{t}$
%\begin{align*}
%\boldsymbol{u}_{t}(x) & =\sum_{i}u_{t}^{i}\boldsymbol{e}_{i},
%\end{align*}
%and $u_{t}^{1}\neq0$ by the regularity assumption on $(t,x)$. 
On the one hand, we have
\[
\boldsymbol{u}_{t}\boldsymbol{\cdot}\boldsymbol{n}=\boldsymbol{u}_{t}\boldsymbol{\cdot}\boldsymbol{e}_1=\det\begin{pmatrix} | & | & & |\\ \boldsymbol{u}_{t} & \boldsymbol{e}_{2} & \cdots & \boldsymbol{e}_{n}\\ | & | & & |\end{pmatrix} = \boldsymbol{u}_{t} \wedge \boldsymbol{e}_{2} \wedge \cdots \wedge \boldsymbol{e}_{n}
= \boldsymbol{u}_{t} \wedge\bigwedge_{i=2}^n\boldsymbol{e}_{i} .% = u_{t}^{1}.
\]
On the other hand, we first observe that
\begin{equation}
\det\left(\mathrm{d}\varphi_{0}^{t}\mathrm{d}\Psi\bigr|_{(t,x)}\right)=\det\mathrm{d}\varphi_{0}^{t}\det\mathrm{d}\Psi\bigr|_{(t,x)}=\det\mathrm{d}\Psi\bigr|_{(t,x)},\label{eq:det}
\end{equation}
since $\det\mathrm{d}\varphi_{0}^{t}=1$ for all $t$. Finally, we compute $\det\left(\mathrm{d}\varphi_{0}^{t}\mathrm{d}\Psi\bigr|_{(t,x)}\right)$
as the change of volume under the action of $\mathrm{d}\varphi_{0}^{t}\mathrm{d}\Psi\bigr|_{(t,x)}$:
%is readily seen that, in the chosen coordinates, $\mathrm{d}\varphi_{0}^{t}\mathrm{d}\Psi$
%takes the form (recall $\partial_t\Psi=\mathrm{d}\Psi\boldsymbol{e}_t=\Psi_*\boldsymbol{e}_t=\boldsymbol{w}$ from Section \ref{sec:coordinates})
\[
\frac{\left(\mathrm{d}\varphi_{0}^{t}\mathrm{d}\Psi\bigr|_{(t,x)}\boldsymbol{e}_{t}\right)\wedge\bigwedge_{i=2}^n\left(\mathrm{d}\varphi_{0}^{t}\mathrm{d}\Psi\bigr|_{(t,x)}\boldsymbol{e}_{i}\right)%\wedge\ldots\wedge\left(\mathrm{d}\varphi_{0}^{t}\mathrm{d}\Psi\bigr|_{(t,x)}\boldsymbol{e}_{n}\right)
}{\boldsymbol{e}_{t}\wedge\boldsymbol{e}_{2}\wedge\ldots\wedge\boldsymbol{e}_{n}}
=\mathrm{d}\varphi_{0}^{t}\boldsymbol{w}_{t}\wedge\bigwedge_{i=2}^n\boldsymbol{e}_{i}%\ldots\wedge\boldsymbol{e}_{n}
= \boldsymbol{u}_{t} \wedge\bigwedge_{i=2}^n \boldsymbol{e}_{i},% \wedge \cdots \wedge \boldsymbol{e}_{n},
\]
%
%
%\mathrm{d}\varphi_{0}^{t}\mathrm{d}\Psi\bigr|_{(t,x)}=\mathrm{d}\varphi_{0}^{t}\begin{pmatrix}| & 0 & \cdots & 0\\
%| & 1\\
%\boldsymbol{w}_t(x) &  & \ddots\\
%| &  &  & 1
%\end{pmatrix}=\begin{pmatrix}\boldsymbol{u}_{t}(x) & \boldsymbol{e}_{2} & \cdots & \boldsymbol{e}_{n}\end{pmatrix},
%\]
where we have used  Eqs.\ \eqref{eq:streak_vf} and \eqref{eq:action}. With Eq.\ \eqref{eq:det},
this finishes the proof.

\section{Proof of \texorpdfstring{$\det\mathrm{d}\Psi=\hat{\boldsymbol{u}}\boldsymbol{\cdot}\hat{\boldsymbol{n}}$}{det(dPsi)=u*n}}
\label{sec:calculation2}

We choose a local unit volume tangent space basis as in Section \ref{sec:calculation}. 
The tangent space $T_{(t,x)}\mathcal{H}$ is then spanned by the orthogonal basis
$\left(\boldsymbol{e}_t+\beta\boldsymbol{e}_1,\boldsymbol{e}_2,\ldots,\boldsymbol{e}_n\right)$.
With $\alpha=\lVert\boldsymbol{e}_t+\beta\boldsymbol{e}_1\rVert=\sqrt{1+\beta^2}$,  
we have that $\frac{1}{\alpha}(\boldsymbol{e}_t+\beta\boldsymbol{e}_1)\wedge\boldsymbol{e}_2\wedge\ldots\wedge\boldsymbol{e}_n$
is a unit $n$-volume parallelepiped in $T_{(t,x)}\mathcal{H}$.

Next, we calculate $\det\left(\mathrm{d}\phi_0^t\mathrm{d}\Psi\right)$ as the change of
volume under $\mathrm{d}\phi_0^t\mathrm{d}\Psi\bigr|_{(t,x)}$:
\begin{align*}
\det\left(\mathrm{d}\phi_0^t\mathrm{d}\Psi\bigr|_{(t,x)}\right) &= \frac{\left(\frac{1}{\alpha}\mathrm{d}\phi_0^t\mathrm{d}\Psi\bigr|_{(t,x)}(\boldsymbol{e}_t+\beta\boldsymbol{e}_1)\right)\wedge\bigwedge_{i=2}^n\mathrm{d}\phi_0^t\mathrm{d}\Psi\bigr|_{(t,x)}\boldsymbol{e}_i}{\frac{1}{\alpha}(\boldsymbol{e}_t+\beta\boldsymbol{e}_1)\wedge\boldsymbol{e}_2\wedge\ldots\wedge\boldsymbol{e}_n}\nonumber\\
&= \tfrac{1}{\alpha}(\boldsymbol{u}_t(x)-\beta\boldsymbol{e}_1)\wedge\boldsymbol{e}_2\wedge\ldots\wedge\boldsymbol{e}_n\nonumber\\
&= \tfrac{1}{\alpha}(\boldsymbol{u}_t(x)-\beta\boldsymbol{e}_1)\boldsymbol{\cdot} \boldsymbol{n}.%\label{eq:calculation}
\end{align*}
From the first to the second line, we have used that the denominator is normalized, 
and the different signs are due to Eq.\ \eqref{eq:action}.

On the other hand, let us calculate the $\hat{\boldsymbol{n}}$-component of the 
extended velocity $\hat{\boldsymbol{u}}_t=\boldsymbol{e}_t+\boldsymbol{u}_t$,
i.e., the component normal to the section's tangent space $T_{(t,x)}\mathcal{H}$:
\begin{align*}
(\boldsymbol{e}_t+\boldsymbol{u}_t)\boldsymbol{\cdot}\hat{\boldsymbol{n}} &= 
(\boldsymbol{e}_t+\boldsymbol{u}_t)\wedge\tfrac{1}{\alpha}(\boldsymbol{e}_t+\beta\boldsymbol{e}_1)\wedge\boldsymbol{e}_2\wedge\ldots\wedge\boldsymbol{e}_n\\
&= \tfrac{1}{\alpha}\boldsymbol{e}_t\wedge(\boldsymbol{u}_t-\beta\boldsymbol{e}_1)\wedge\boldsymbol{e}_2\wedge\ldots\wedge\boldsymbol{e}_n\\
&= \tfrac{1}{\alpha}(\boldsymbol{u}_t-\beta\boldsymbol{e}_1)\wedge\boldsymbol{e}_2\wedge\ldots\wedge\boldsymbol{e}_n\\
&= \tfrac{1}{\alpha}(\boldsymbol{u}_t(x)-\beta\boldsymbol{e}_1)\boldsymbol{\cdot} \boldsymbol{n}.
\end{align*}
From the first to the second line, we have used determinant rules, and from the second to the third
lines Laplace expansion along the time component.

Finally, the above calculation is, of course, consistent with the one given in Section \ref{sec:calculation}, 
since $\boldsymbol{u}\boldsymbol{\cdot}\boldsymbol{n}=\hat{\boldsymbol{u}}\boldsymbol{\cdot}\hat{\boldsymbol{n}}$
whenever $\hat{\boldsymbol{n}}=\left(\begin{smallmatrix} 0 \\ \boldsymbol{n}\end{smallmatrix}\right)$, 
i.e., when the section does not move. In this case, $\beta=0$ and $\alpha=1$.

\section*{Acknowledgements}
I would like to thank Simon Eugster, Florian Huhn and Dietmar Salamon
for useful discussions, as well as George Haller for helpful comments
on an earlier version of the manuscript. David Legland's \texttt{geom2d}-library
for \textsc{Matlab} was very useful in the loop decomposition used in Fig.\ \ref{fig:donating_region}.

\bibliographystyle{plainurl}

\end{document}